\documentstyle[12pt]{article}
\begin{document}
\pagestyle{empty}

\input epsf
\font\ssqfont=cmssq8 scaled 2500
\font\sqfont=cmssq8 scaled 1100
\null\vspace*{-104pt}
\begin{flushright}
{\footnotesize
FERMILAB--PUB--97/068--A\\
astro-ph/9704070\\
April 1997}
\end{flushright}

\begin{center}
\vspace{.2in}
{\Large \bf Ribbons on the CBR Sky:  A Powerful Test of \\
\vspace{12pt}
A Baryon Symmetric Universe}
\bigskip

\vspace{.2in}
William H. Kinney,$^1$ Edward W. Kolb,$^{1,2}$
and Michael S. Turner$^{1,2,3}$\\

\vspace{0.2in}
{\it $^1$NASA/Fermilab Astrophysics Center\\
Fermi National Accelerator Laboratory, Batavia, IL~~60510-0500}\\

\vspace{.1in}
{\it $^2$Department of Astronomy \& Astrophysics\\
Enrico Fermi Institute, The University of Chicago, Chicago, IL~~60637-1433}\\

\vspace{0.1in}
{\it $^3$Department of Physics\\
The University of Chicago, Chicago, IL~~60637-1433}\\

\end{center}

\vspace{.2in}

\baselineskip=24pt

\begin{quote}
If the Universe consists of domains of matter and antimatter,
annihilations at domain interfaces leave a distinctive imprint on the
Cosmic Background Radiation (CBR) sky.  The signature
is anisotropies in the form of long, thin ribbons of width $\theta_W\sim
0.1^\circ $, separated by angle $\theta_L\simeq
1^\circ(L/100h^{-1}\mbox{Mpc})$ where $L$ is the characteristic domain
size, and $y$-distortion parameter $y \approx 10^{-6}$.
Such a pattern could potentially be detected by the high-resolution CBR
anisotropy experiments planned for the next decade, and
such experiments may finally settle the question of whether
or not our Hubble volume is baryon symmetric.
\vspace*{12pt}

\end{quote}

\newpage
\pagestyle{plain}
\setcounter{page}{1}

The conventional view is that the Universe possesses a {\it baryon
asymmetry,} and all astrophysical objects are made of baryons.  This is
quite a reasonable view.  Clearly there is a local asymmetry between
matter and antimatter: Earth is made entirely of matter, as well as
the Moon, as evidenced by the fact that Apollo astronauts took a
second small step.  On scales beyond the solar system the arguments
become less direct and less compelling.  About the strongest
statement one can make is that if the Universe is baryon symmetric, matter and
antimatter must be separated into domains at least as large as
the size of clusters of galaxies, $L \sim 20\,{\rm Mpc}$ \cite{steigman76}.

Although the simplest picture is that the Universe possesses a {\it
global} baryon asymmetry, the possibility of a symmetric Universe in
which matter and antimatter are separated into very large domains of
equal, but opposite, baryon number has been discussed
over the years \cite{stecker81}.
As de\,Rujula has recently emphasized, even if matter and antimatter
are segregated on very large scales, $L\sim 20\,$Mpc, it may be
possible to detect the presence of antimatter \cite{derujula96}.  One
direct approach is to search for antinuclei in cosmic rays
\cite{ahlen94}.  Another is to look for the products
of matter--antimatter annihilations from domain boundaries, e.g.,
high-energy gamma rays \cite{derujula96}.
A third possibility, which is the subject of this
paper, is to look for a signature of matter--antimatter
annihilations as distortions in the Cosmic Background Radiation (CBR).
As we shall describe, the signature is very robust
as the physics is straightforward, and further, it allows scales
as large as the Hubble length ($\sim 3000 \,$Mpc) to be probed.

Heat is generated at the domain interfaces due to nucleon--antinucleon
($N$--$\overline{N}$) annihilations.  Around the time of last
scattering of the background photons\footnote{Throughout the paper
``last scattering'' refers to the epoch of last scattering of CBR photons,
and will be abbreviated ``LS.''  We assume standard recombination
so that $z_{\rm LS} \simeq 1100$; measurements of CBR anisotropy
on angular scales of around $1^\circ$ make a very strong case
for standard recombination \cite{stdrecombination}.}
the injected energy cannot be
thermalized, and it distorts the Planckian spectrum of the CBR.
The spatial pattern of distortions is ribbon-like linear
structures with angular width characterized by the photon
diffusion length at recombination, $\theta_W\simeq 0.1^\circ$, and
separation that depends on the domain size, $\theta_L \simeq 1^\circ
(L/100h^{-1}\,$Mpc); see Fig.~1.  The CBR
distortion caused by $N$--$\overline{N}$ annihilations takes the form
of a Sunyaev--Zel'dovich $y$ distortion \cite{zeldovich69} with
magnitude $y \simeq 10^{-6}$.  A $y$ distortion corresponds to
a frequency-dependent temperature fluctuation
\cite{zeldovich69}
\begin{equation}
{\delta T(\nu )\over T}  = y\left[ \left( \frac{\nu}{\nu_0} \right)
		\frac{\exp(\nu/\nu_0)+1}{\exp(\nu/\nu_0) -1} -4\right]
\longrightarrow \left\{   \begin{array}{rl}
-2y & \nu \ll \nu_0 \\
y(\nu/\nu_0) & \nu \gg \nu_0
\end{array}\right. ,
\end{equation}
where $\nu_0=kT/h=56.8\,$GHz.  At low frequencies the
$y$ distortion is independent of $\nu$, and hence indistinguishable
from a true temperature fluctuation of magnitude $\delta T/T = -2y$.

\begin{figure}
\centerline{\Large $L=600h^{-1}$Mpc \hspace*{1.0in}  $L=1200h^{-1}$Mpc}
\hspace*{25pt} \epsfxsize=200pt \epsfbox{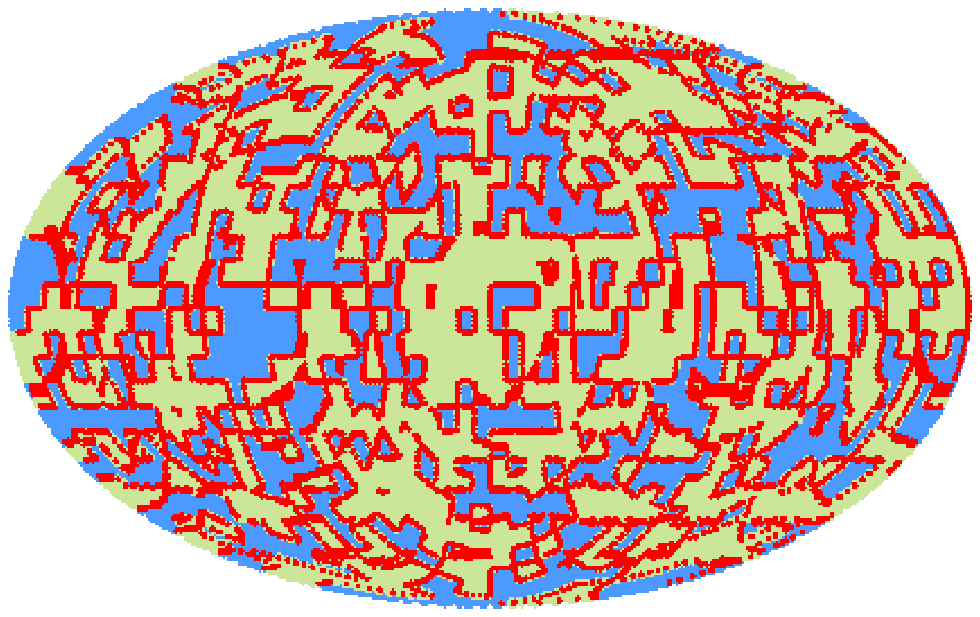}
\epsfxsize=200pt \epsfbox{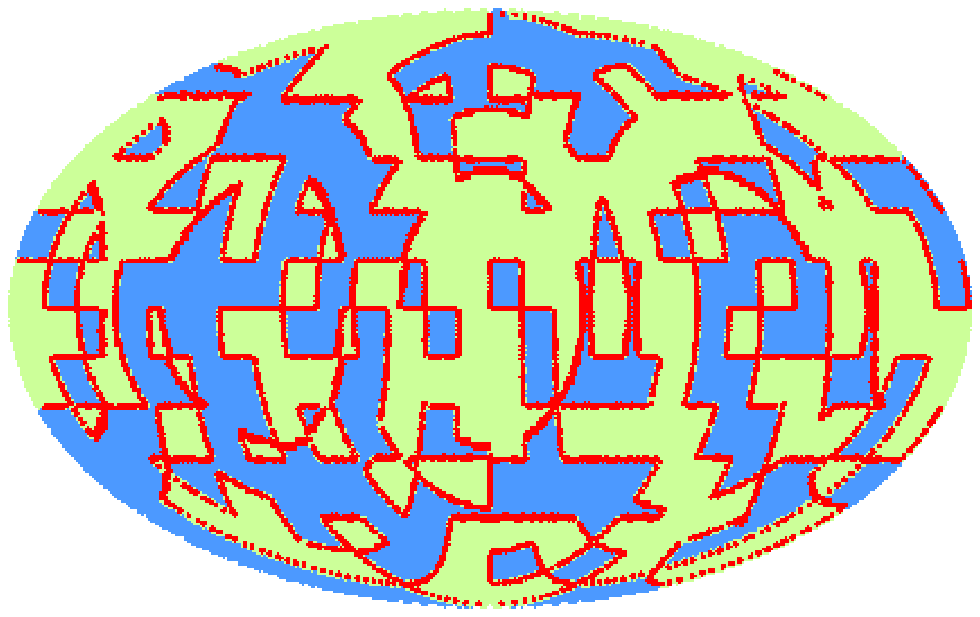}\\
\vspace{12pt}
\centerline{\Large $L=2400h^{-1}$Mpc \hspace*{1.0in}  $L=6000h^{-1}$Mpc}
 \hspace*{25pt}\epsfxsize=200pt \epsfbox{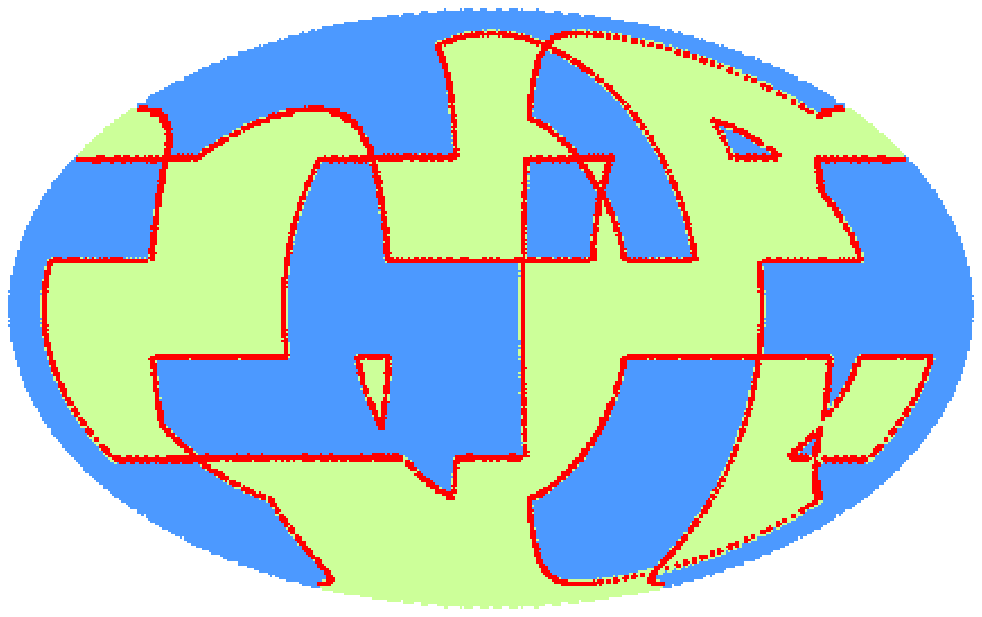}
\epsfxsize=200pt \epsfbox{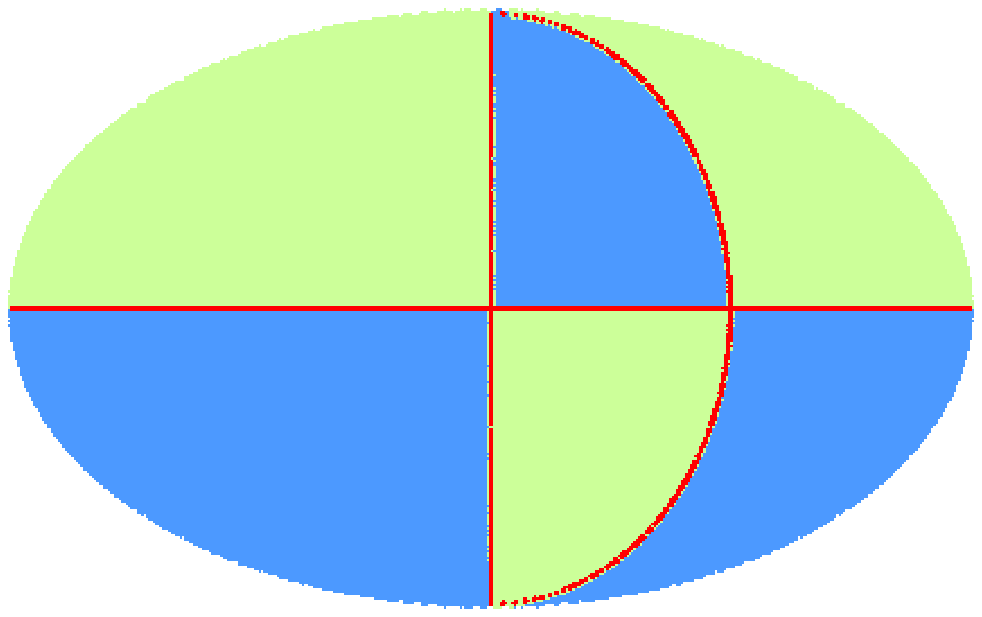}\\
\hspace*{1em} {\footnotesize{{\bf Fig.\ 1:}  Realizations of
matter -- antimatter distortions on the CBR sky for various
cubic domain sizes $L$.  Domains of opposite baryon asymmetry are
shown in contrasting shadings, with the interfaces between domains
highlighted.   The spectral distortion of the CBR is confined to
these interfaces, which appear as long ``ribbons.'' }}
\end{figure}

The pattern and the amplitude of CBR anisotropy from
$N$--$\overline{N}$ annihilations is interesting because it is not
excluded by the present generation of CBR experiments, but should be
within the range of the next round of large-area, high-resolution
experiments (e.g., NASA's MAP satellite and ESA's Planck).

To orient the reader, we begin with a rough estimate of the $y$
distortion, and then proceed with a more careful calculation.  In the
discussion below, $h\equiv H_0/100\,{\rm km\, sec^{-1}\,Mpc^{-1}}$ and
the baryon density is quantified by $\Omega_Bh^2$.  
We take as representative values $h=1/2$ and
$\Omega_Bh^2 =0.024$, the latter based upon recent
determinations of the primeval deuterium abundance in high-redshift
hydrogen clouds \cite{tytler}.

We assume that some process in the early Universe produced regions of
equal and opposite baryon number \cite{example}, with $|n_B-n_{\bar
B}|/n_\gamma \equiv \eta = 6.5 \times 10^{-10} (\Omega_B
h^2/0.024)$.\footnote{In Ref.\ \cite{derujula96}, de\,Rujula has
argued on the basis of the uniformity of the CBR sky on large angular
scales that the magnitude of the baryon asymmetry must be nearly
identical in matter and antimatter domains.}  If we divide the
Universe into cells of comoving size $L$ populated equally with matter
and antimatter, then individual cells will be part of larger clusters
in a percolation pattern.  Interfaces separating matter domains and
antimatter domains will have a surface area $A$ that is much larger
than $L^2$.  The magnitude of the $y$ distortion does not depend
upon $L$ or $A$.

Consider matter--antimatter annihilations occurring in the
interface regions.  Because the electron mass is so much smaller than
the nucleon mass, the heat released is dominated by
$N$--$\overline{N}$ annihilations.  Nucleon rest-mass energy is
released through the production and subsequent decay of pions
\cite{steigman76}:
\begin{equation}
N + \overline{N} \rightarrow
\left\{
\begin{array}{l}
 \pi^0 \rightarrow \gamma + \gamma  \\
 \pi^\pm \rightarrow \mu^\pm +
\nu_\mu\left(\bar\nu_\mu\right)   \\
\hspace*{36pt}
\begin{picture}(30,12)\put(8,12){\oval(10,18)[bl]}\put(8,3){\vector(3,0){14}}
\end{picture}
e^\pm + \nu_e\left(\bar\nu_e\right) +  \bar\nu_\mu\left(\nu_\mu\right).
\end{array}
\right.
\end{equation}
Half the total annihilation energy is in the form of neutrinos,
one-third is in the form of $\langle E \rangle \sim 200\,{\rm MeV}$
photons, and one-sixth is in the form of $\langle E \rangle\sim
100\,{\rm MeV}$ electrons and positrons.
Because neutrinos interact only through weak interactions they deposit
negligible energy in the photon gas.  It is also easy to see that
$200\,{\rm MeV}$ photons do not significantly heat the photon gas,
since at the time of last scattering the mean free path of a
$200\,{\rm MeV}$ photon is larger than the Hubble length.

Significant heating comes only from the $100\,{\rm MeV}$ electrons and
positrons.  The scattering of these particles off background photons
is much more efficient than scattering of high-energy photons off
background electrons because there are roughly $10^{10}$ background
photons for every background electron.  The $100\,{\rm MeV}$ electrons
and positrons quickly lose their energy to background
photons via inverse Compton scattering, and the upscattered photons
slowly lose energy and heat the CBR photons producing the $y$ distortion.
As a first approximation, we assume
that all the energy carried by 100\,MeV electrons and positrons heats
the photon gas.  This means that the total energy dumped into the CBR
per $N$-$\overline{N}$ annihilation is $2m_N/6$, where $m_N$
is the nucleon mass.

The $N$-$\overline{N}$ annihilation cross section is so large that
well after last scattering any nucleon (antinucleon) that drifts into
an antimatter (matter) domain is annihilated on a timescale much less
than a Hubble time.  The transverse thickness of the annihilation
region is proportional to the nucleon free streaming distance at the
time of last scattering, approximately $v_{\rm LS} H^{-1}_{\rm
LS}$. Here, $v_{\rm LS}$ is the nucleon velocity dispersion at the
time of last scattering, $v^2_{\rm LS} = 3T_{\rm LS}/m_N$. Expressed
as a comoving length, $\lambda_{\rm FS}(R_{\rm LS} )\simeq
v_{\rm LS} H^{-1}_{\rm LS}R_{\rm LS}^{-1} = 5 \times 10^{-3}
(0.5/h)\,{\rm Mpc}$, where $R$ is the cosmic scale factor, normalized
to unity today with $R_{\rm LS} = 9.1\times10^{-4}$.  A better
approximation for the thickness is $2 \lambda_{\rm FS}(R_{\rm LS} ) /
\sqrt{3}$, where the factor of $2$ comes from the fact that nucleons
diffuse into antimatter regions and antinucleons diffuse into matter
regions, and the factor $1/\sqrt{3}$ is the projection of the velocity
in the transverse direction.

The number density of annihilation pairs is $\eta n_\gamma/2$, and the
amount of energy released per annihilation is $2m_N/6$.  The amount of
heat produced per cross-sectional area $A$ perpendicular to the
interface region is
\begin{equation}
\frac{\Delta Q}{A} = \frac{2\lambda_{\rm FS}(R_{\rm LS} )}{\sqrt{3}}
	\, \frac{\eta n_\gamma}{2} \frac{2m_N }{6}\ .
\label{rough1}
\end{equation}

By the time of last scattering the heat deposited by 100\,MeV
electrons and positrons in the interface region will spread into a
larger region.  The thickness of this region is governed by photon
diffusion around last scattering, and the relevant length scale is the
Silk scale, $\lambda_S \simeq 22 \,(0.012/\Omega_Bh^3)^{1/2}\,{\rm
Mpc}$, again expressed as a comoving length \cite{silk68,hu95}.

The fractional increase in the energy of the photons in the photon
diffusion region, $\Delta Q/Q$, determines the magnitude of the CBR
anisotropy.  Since $\Delta Q/A$ is spread out over a thickness
$2\lambda_S/\sqrt{3}$ (the factors of $2$ and $\sqrt{3}$ arise from
the considerations discussed above) and the heat energy in photons
per area at last scattering is $Q/A= 2.7 T_{\rm LS} n_{\gamma}
\left(2\lambda_S/\sqrt{3}\right)$, the fractional change is
\begin{equation}
\label{rough2}
{\Delta Q \over Q} \simeq
\frac{2\lambda_{\rm FS}(R_{\rm LS} )/\sqrt{3}}{ 2\lambda_S/\sqrt{3}}\
\eta \ {m_N/6 \over 2.7 T_{\rm LS}}\,
 =  3.2\times 10^{-5}\, \left(\frac{\Omega_Bh^2}{0.024}\right)^{3/2}
\left(\frac{h}{0.5}\right)^{-1/2}.
\end{equation}
Since the heat deposited in the annihilation region,
which is larger by a factor of $\lambda_S/\lambda_{\rm FS} \sim
10^4$, is a small perturbation, any backreaction on the
annihilation process itself can be safely ignored.

As mentioned earlier, energy from $N$--$\overline{N}$ annihilation
leads to a $y$ distortion, with magnitude $y = {1\over 4} \Delta Q/Q$.
For frequencies much less than about 100\,GHz, the $y$ distortion is
indistinguishable from a temperature anisotropy of magnitude
\begin{equation}
{\delta T\over T} = -2y = -\frac{1}{2}\, \frac{\Delta Q}{Q}
  \simeq -1.6 \times 10^{-5}\,  \left(\frac{\Omega_Bh^2}{0.024}\right)^{3/2}
\left(\frac{h}{0.5}\right)^{-1/2}.
\end{equation}
Note that at low frequencies the ribbons
appear {\it cooler} than the surrounding,
unheated regions of the CBR sky.
The width of the photon diffusion region determines
the angular width of the ribbons,
\begin{equation}
\theta_W \simeq \frac{2 \lambda_S/\sqrt{3}}{2H_0^{-1}} \simeq 0.1^\circ\,
\left(\frac{0.05}{\Omega_B h}\right)^{1/2}.
\end{equation}
The CBR anisotropy from $N$--$\overline{N}$ annihilations should
take the form of linear features, or ``ribbons,'' of width
$0.1^\circ$ and characteristic separation
$\theta_L \simeq 1^\circ (L/100h^{-1}\,$Mpc) set by the domain size.
The spatial pattern is illustrated in Fig.\ 1.

This rough estimate neglects some potentially important effects:
the efficiency with which 100\,MeV electrons and positrons
from annihilations heat the ambient photons, the fact that some heating
occurs before last scattering, the expansion of the Universe, and,
most importantly, the fact that the diffusion length of
protons and antiprotons is much smaller than the free streaming
length $\lambda_{\rm FS}$ due to Coulomb scattering.
We now refine our calculation; the net result is a reduction
in the estimate for $y$ by about a factor of ten.

To begin, the most important nucleons are those in
neutral atoms, hydrogen, antihydrogen, helium and antihelium,
because their free streaming is not inhibited by Coulomb scattering.
Hydrogen formation occurs at a redshift $z_{\rm H-REC} \sim 1500$
and helium formation occurs slightly earlier,
at a redshift $z_{\rm HE-REC} \sim 2800$. We assume that recombination 
is instantaneous, which is a better approximation for helium than for 
hydrogen.

Next, let's follow the energy flow from annihilations more carefully.
1) One-sixth of the annihilation energy goes into 100\,MeV
electrons and positrons. 2) The 100\,MeV electrons and positrons
quickly lose energy via inverse Compton scattering off background
photons, producing photons of typical energy $E_\gamma \simeq
3\gamma^2T = 1.2 \times 10^5 T = 2.8\times 10^{-5}R^{-1}$\,MeV, where
$\gamma = E_e/m_e \simeq 200$.  We are interested in the
interval between the equality of radiation and matter energy densities
($R_{\rm EQ} = 4.2\times10^{-4}h^{-2}$) and last scattering
($R_{\rm LS}=9\times10^{-4}$), so $E_\gamma$ is in the range
0.03 to 0.15\,MeV.  We refer to the photons produced in this step
as ``secondary'' photons.  3) The secondary photons slowly lose energy
by Thomson scattering off ambient electrons (with energy loss of about
$ E_\gamma^2/m_e$ per scattering).  4) Finally, the electrons produced
in the third step rapidly lose energy to the background photons.  The
last step is the means by which the $y$ distortion arises; the
penultimate step is the rate limiting step.

We refine Eq.\ (\ref{rough2}) by integrating over the interval
between recombination (``REC'') and last scattering (``LS'')
for hydrogen and antihydrogen ($i=$H) and helium and
antihelium ($i=$He) separately:
\begin{equation}
\label{correct}
\left( {\Delta Q\over Q}\right)_i  =
    \int_{\rm REC}^{\rm LS} dR\,  \frac{d\lambda_{\rm FS}(R)}{dR} \,
\frac{\eta}{\lambda_S}\,  {X_i}\, \left( N_\gamma(R)
\frac{ \Delta E_\gamma(R;R_{\rm LS})}{2.7T_{\rm LS}} \right) \ .
\end{equation}
The factor $X_i$ accounts for the mass fraction in hydrogen 
(antihydrogen), about 75\%, and in helium (antihelium), about 25\%.
The factor $d\lambda_{\rm FS}(R)$
accounts for the growth of the annihilation interface. Prior to recombination,
the atoms can be taken to be in thermal equilibrium, with velocity 
$v \propto R^{-1/2}$. Once the atoms recombine, however, they free stream
with a velocity which redshifts as $v \propto R^{-1}$. 
The growth of the annihilation interface is then given by
\begin{equation}
d\lambda_{\rm FS} = \frac{v\left(t\right)}{R\left(t\right)} dt = 
\sqrt{\frac{R_{\rm REC}}{R}} \left(\frac{v_{\rm LS} 
H_{\rm LS}^{-1}}{R_{\rm LS}}\right) d\ln R,
\end{equation}
where $v_{\rm LS} = \sqrt{T_{\rm LS} / M}$ is the thermal velocity at 
last scattering, half as large for helium as for hydrogen.

The term $N_\gamma(R)\Delta E_\gamma (R;R_{\rm LS})$
is the nucleon rest-mass energy liberated into secondary photons
when the scale factor was $R$ and transferred to background photons
by the time of last scattering.  Here, $N_\gamma(R) = (m_N/6)/E_\gamma(R)
\simeq 5.7\times 10^6R$ is the number of secondary photons per nucleon
annihilated and $\Delta E_\gamma(R;R_{\rm LS})$ is the energy
transferred to the background photons by the time of last
scattering by a single secondary photon.

In the absence of interactions, the energy of a secondary photon would
simply scale inversely with the scale factor, and a secondary photon
produced when the scale factor was $R$ would have energy at last
scattering of $ (R/R_{\rm LS})E_\gamma(R)$.  But because the secondary
photon loses energy by scattering, its actual energy at last
scattering, $E_\gamma(R_{\rm LS})$, is less.  The energy
transferred to the background photons by last scattering is this
difference, $\Delta E_\gamma(R;R_{\rm LS}) = (R/R_{\rm LS})E_\gamma(R)
- E_\gamma(R_{\rm LS})$.  In the approximation used previously the
energy transfer was taken to be 100\% efficient ($E_\gamma(R_{\rm
LS})=0$) and instantaneous at last scattering ($R=R_{\rm LS}$), so
$\Delta E_\gamma(R;R_{\rm LS}) = E_\gamma(R=R_{\rm LS})$.  Combined
with the expression for $N_\gamma(R)$, $N_\gamma(R)\Delta E(R;R_{\rm
LS})$ was simply $m_N/6$, and together with the assumption that
everything occurs at last scattering led to Eq.\ (\ref{rough2}).

Now we turn to the calculation of the actual energy of the secondary
photon at $R_{\rm LS}$. The evolution of the energy of the secondary
photon is determined by two effects, a redshift term and a term due to
the transfer of energy to the background electrons (which is then
rapidly transferred to the background photons):
\begin{equation}
dE_\gamma = - E_\gamma \frac{dR}{R} - \frac{E_\gamma^2}{m_e} n_e \sigma_T
dt \ ,
\end{equation}
where $\sigma_T = 6.7\times10^{-25}$cm$^2$ is the Thomson cross section
and the factor $E_\gamma^2/m_e$ is the energy loss suffered by a secondary
photon in Thomson scattering.  This equation can be integrated,
\begin{equation}
\label{yield}
\frac{1}{R_{\rm LS}E_\gamma(R_{\rm LS})} = \frac{1}{RE_\gamma} +
a \left( \frac{1}{R^{5/2}} - \frac{1}{R_{\rm LS}^{5/2}} \right) \ ,
\end{equation}
where $a = {2\over 5} (n_e \sigma_T /H_0 m_e) = 2.7
\times10^{-3} (\Omega_Bh/0.05)$\,MeV$^{-1}$, where $n_e$ is
the present density of electrons.  We can use this
expression to obtain some idea of the efficiency of energy loss of
secondary photons.  Setting $\Omega_Bh/0.05=1$, the two terms on the
right-hand-side are equal for $R=R_{\rm LS}/1.2$, which implies that
a secondary photon will lose more than half of its energy by last
scattering if it is produced at $R<R_{\rm LS}/1.2$.

Using the result of Eq.\ (\ref{yield}) gives
\begin{equation}
\frac{\Delta E_\gamma(R;R_{\rm LS})}{2.7T_{\rm LS}} =
\frac{ 3.3\times 10^{-3}(\Omega_Bh/0.05)(R^{-5/2}-R_{\rm LS}^{-5/2})}
{1 + 7.5 \times 10^{-8}(\Omega_Bh/0.05)(R^{-5/2} - R_{\rm LS}^{-5/2}) } \, .
\end{equation}
All the pieces are now in place to integrate Eq. (\ref{correct});
the result can be given as a dimensionless correction factor
which multiplies our earlier estimate in Eq.~(\ref{rough2}),
\begin{eqnarray}
\left( {\Delta Q\over Q}\right)_i & = & 3.2\times 10^{-5}\,
        \left(\frac{\Omega_Bh^2}{0.024}\right)^{3/2}
        \left(\frac{h}{0.5}\right)^{-1/2}\, C_i \nonumber\\
C_{\rm H} & = & 1.8\times 10^{-2}\, \int_{\rm REC}^{\rm LS}
        dR\, \left(\frac{R_{\rm REC}}{R}\right)^{1/2}
        \frac{ 3.3\times 10^{-3}(\Omega_Bh/0.05)
        (R^{-5/2}-R_{\rm LS}^{-5/2})}
        {1 + 7.5\times10^{-8}(\Omega_Bh/0.05)(R^{-5/2} - R_{\rm LS}^{-5/2}) }
        \nonumber \\
& \simeq & 0.1 (\Omega_B h/0.05)^{0.4} \nonumber \\
C_{\rm He} & = & 3.0\times 10^{-3}\, \int_{\rm REC}^{\rm LS}
        dR\,  \left(\frac{R_{\rm REC}}{R}\right)^{1/2}
        \frac{ 3.3\times 10^{-3}(\Omega_Bh/0.05)
        (R^{-5/2}-R_{\rm LS}^{-5/2})}
        {1 + 7.5\times10^{-8}(\Omega_Bh/0.05)(R^{-5/2} - R_{\rm LS}^{-5/2}) }
        \nonumber \\
& \simeq & 0.04 (\Omega_B h/0.05)^{0.2}
\end{eqnarray}
where the final expressions are numerical fits.
Putting it all together, our final result for the distortion parameter is
\begin{equation}
y =
\frac{1}{4}\sum_i \left( \frac{\Delta Q}{Q}\right)_i \simeq
 10^{-6} (\Omega_B h^2/0.024)^{1.9} (h / 0.5)^{-1/2} \ ,
\end{equation}
which indicates that our rough estimate was about a factor of ten too high.

While this result is based upon a more careful calculation,
it should still be regarded as an estimate.  For example, the
diffusion of the annihilation heat was approximated by the
characteristic scale $\lambda_S$; a more careful treatment would
properly treat diffusion, the visibility function for last
scattering, geometric effects, and the details of recombination.  
There is another $y$ distortion
with a less distinctive signature that arises from annihilation
surfaces along the line-of-sight between here and last scattering.
This leads to a $y$ distortion which is proportional to $1/L$ and
which covers the CBR sky like a blanket.  This distortion was
first discussed in the context of a well mixed, baryon-symmetric
Universe by Sunyaev and Zel'dovich \cite{blanket}. (Jones and Steigman
also discussed $y$ distrortions in a variety of scenarios\cite{jones78}.)  
We will address all of these issues in a future paper.

In conclusion, if large domains of matter and antimatter are present
in the Universe,\footnote{Note that our analysis does not require
equal number of matter and antimatter domains, so long as both are
abundant enough to percolate and form large regions.} energy released
from annihilation at their boundaries around the time of last
scattering produces a distinct signature on the CBR sky: A
Sunyaev-Zel'dovich $y$ distortion of magnitude $10^{-6}$ in
the form of thin ribbons on the sky with width $0.1^\circ$
and separation determined by the domain size $L$,
$\theta_L \simeq 1^\circ (L/100h^{-1}\,{\rm Mpc})$.

The ribbon feature should be detectable by the high-resolution, full-sky
anisotropy maps that will be produced by NASA's MAP mission and ESA's
Planck mission, or perhaps earlier by earth-based and balloon-borne
experiments with better than sub-degree angular resolution and large
sky coverage (e.g., VCA, VSA, Boomerang or TopHat).
Because the CBR sky allows us to probe scales as large as
the Hubble length, CBR experiments have the potential to settle the question
of the matter/antimatter composition of the observable Universe.

\paragraph{Acknowledgments.}  We thank Albert Stebbins and
Scott Dodelson for valuable
discussions and acknowledge useful communications with Alvaro De Rujula.
This work was supported by the DoE (at Chicago and Fermilab)
and by the NASA (at Fermilab by grant NAG 5-2788).


\frenchspacing

\begin{thebibliography} {cst}

\bibitem{steigman76} G. Steigman, {\it Ann. Rev. Astron. Astrophys.}
{\bf 14}, 339 (1976).

\bibitem{stecker81} F. W.  Stecker,  {\it Ann. NY Acad. Sci.}
{\bf 375}, 69 (1981)

\bibitem{derujula96} A. De Rujula, New Ideas for Dark Matter,
in {\em Proceedings of the 7th International Meeting on Neutrino Telescopes,}
edited by M. Baldo-Ceolin.

\bibitem{ahlen94} S. P. Ahlen {\it et al}, {\it N.I.M.} {\bf A350}, 351 (1994).

\bibitem{stdrecombination} See e.g., S.~Dodelson, astro-ph/9702134.

\bibitem{zeldovich69} See e.g., R. A. Sunyaev and Ya. B. Zel'dovich,
{\it Ann. Rev. Astron. Astrophys.} {\bf 18}, 537 (1980).

\bibitem{tytler} D.~Tytler, X.-M.~Fan and S.~Burles, {\it Nature} {\bf 381},
207 (1996); D.~Tytler, S.~Burles and D.~Kirkman, astro-ph/9612121.

\bibitem{example} See e.g., V. A. Kuzmin, M.E. Shaposhnikov and
I. I. Tkachev, {\it Phys. Lett. B} {\bf 105}, 167 (1981).

\bibitem{silk68} J. Silk, {\it Astrophys. J.} {\bf 151}, 459 (1968).

\bibitem{hu95} W. Hu and N. Sugiyama, {\it Astrophys. J.} {\bf 444}, 489
(1995).

\bibitem{blanket} R.~Sunyaev and Ya.B.~Zel'dovich, {\it Astrophys. Sp. Sci.}
{\bf 7}, 20 (1974); also see Ref.~\cite{steigman76}, pp. 364-366.

\bibitem{jones78} B. J. T. Jones and G. Steigman, {\it Mon. Not. R. Astron. 
Soc.} {\bf 183}, 585 (1978).

\end{thebibliography}
\end{document}